\documentstyle[12pt]{article}
\newcommand{\bibi}{\bibitem}

\newcommand{\prl}{\it Phys. Rev. Lett.}

\newcommand{\prb}{\it Phys. Rev. B}

\newcommand{\half}{\frac {1}{2}}

\newcommand{\beq}{\begin{equation}}
\newcommand{\eeq}{\end{equation}\noindent}

\newcommand{\beqr}{\begin{eqnarray}}
\newcommand{\eeqr}{\end{eqnarray}\noindent}

\begin{document}
\title {Confinement in the Three-dimensional Anisotropic $t$-$J$
 Model: A Mean-Field Study}

\author{Sanjoy K. Sarker \\
Department of Physics and Astronomy \\
The University of Alabama \\
Tuscaloosa, Alabama 35487 \\}
\date{}
\maketitle
\begin{abstract}
We consider the anisotropic $t$-$J$ model with the c-axis parameters
$t_c$ and $J_c$ different from their in-plane counterparts, $t$ and $J$.
Within the slave-fermion mean-field approximation it is shown that
the spiral state exhibits charge-confinement in the intermediate $\delta$
regime for a range of values of $t_c/t$.
In the confined state the hopping amplitude
$<c^{\dag}_{i\sigma}c_{j\sigma}> = 0$  along the c direction
so that c-axis resitivity is infinite at $T = 0$.

\end{abstract}
\vspace{0.1in}
\pagebreak

The normal-state resistivity of the cuprate superconductors is
metallic in the $ab$ plane but is characteristic of an insulator
along the $c$ direction, leading to the possibility of the remarkable
phenomenon of confinement \cite{and,cla}.
By continuity, such a behavior is not expected to occur in a Fermi
liquid. In addition, confinement is at the heart of
the proposed pair-tunneling mechanism which is
in essence a deconfining process \cite{chak}.
Theoretical treatments so far have been focused on a collection
of weakly coupled Hubbard chains \cite{cla}.
In this paper we study confinement in a collection of Hubbard planes,
more specifically
in the anisotropic $t$-$J$ model described by the Hamiltonian
\beq H = - \sum _{ij} t_{ij}c^{\dag}_{i\sigma}c_{j\sigma}
+  \frac{1}{2} \sum _{ij} J_{ij}\lbrack S_i\cdot S_j -
n_in_j \rbrack. \eeq
Here the in-plane hopping parameter $t_{ab}  \equiv t$ and the exchange
interaction $J_{ab} \equiv J$ are in general different $t_c$ and $J_c$,
the corresponding quantities along the $c$ axis. Not all the
parameters are free since the $t$-$J$
model is thought to be derived from an underlying Hubbard model
($J_{ij} = 4t^2_{ij}/U$) so that $J_c/J = (t_c/t)^2$. There are
therefore three independent parameters: $t/J$,
$\xi \equiv t_c/t$ and the hole density $\delta$.

Confinement is presumably intimately connected with spin-charge
separation. While model (1) is not exactly solvable, in two
dimensions a number of approximate ground states have been proposed
that exhibit spin-charge separation. Here we study the spiral
states in the Schwinger-boson slave-fermion representation:
$c_{i\sigma} = h^{\dag}_ib_{i\sigma}$, where $h^{\dag}$ creats a
fermionic hole and $b_{i\sigma}$ destroys a bosonic spin \cite{jay}.
We will impose the constraint
$h^{\dag}_ih_i + \sum _{\sigma}b^{\dag}_{i\sigma}b_{i\sigma} = 1$
on the average. A mean-field decomposition leads to following
hamiltonians:
\beq H_h = 2 \sum _{ij} t_{ij}B_{ij} h^{\dag}_ih_j \eeq
\beq H_b =  \sum _{ij\sigma} t_{ij}D_{ij}b^{\dag}_{i\sigma}b_{j\sigma}
  -  \sum _{ij\sigma} J_{ij}A_{ij}\sigma b_{i\sigma}b_{j-\sigma}.\eeq
Here $D_{ij} = <h^{\dag}_ih_j>$ is the average hopping amplitude. This
is associated with ferromagnetic backflow
$B_{ij} = <b^{\dag}_{i\sigma}b_{j\sigma}>$. And
$A_{ij} = \half <(b_{i\uparrow}b_{j\downarrow} -
b_{i\downarrow}b_{j\uparrow})>$ represents the antiferromagnetic
correlations associated with the exchange term. We have shown that
in two dimensions this competition gives rise to \cite{jay,sar}
an incommensurate spiral metallic state that
evolves continuously from the Neel state at
$\delta = 0$ to a ferromagnetic state at large $t\delta/J$.
The spiral state is favored over double spiral and canted states
and is stabilized against phase separation and domain walls by
Coulomb repulsion \cite{sar,hu}.

In the present problem the in-plane amplitudes ($D_{ab},
B_{ab}$ and $A_{ab}$) are in general different from
those along the $c$ directions ($D_c, B_c$ and $A_c$).
We have solved the mean-field equations numerically for various
values of $t_{ab}/J_{ab}, \xi = t_c/t_{ab}$ and $\delta$. One
self-consistent solution is found to be
the usual spiral metallic state which evolves continuously out of
the Neel state. For this state all the mean-field amplitudes
including $D_c$ and $B_c$ are nonzero and hence there is no
confinement. In addition, the state with $D_c = 0$ (and hence
$B_c = 0$) is also a self-consistent solution. This is clearly
a direct consequence of spin-charge separation. Charge can propagate
in the $ab$ plane in this case ($D_{ab} \ne 0$).
Hence $D_c$ can be thought of
as the order-parameter for deconfining transitions.

While spin-charge separation can lead to confinement, it does not
guarantee that such a state is favored energetically.
The ground-state energy is given by
\beq E_G = 8tD_{ab}B_{ab} - 4JR_{ab} + 4t_cD_cB_c
- 2J_cR_c, \eeq
where $R_{ij} = A^2_{ij} - B^2_{ij}/2 + (1-\delta)^2/8$.
In general $D$ and $B$ have opposite signs and $R > 0$. For
physically interesting values of $t-J$, we find that the
unconfined phase has a lower energy both at small and large
$\delta$. But, as shown in Fig. 1, for intermediate $\delta$ there
is a region in the $\delta $-$t_c/t$ plane where the {\em confined}
state is favored. Interestingly, for fixed $t/J$ and $\delta$, the
mean-field
state is unconfined at small $t_c/t$, and with increasing $t_c/t$
there is a first-order transition to a confined state.  This is
because as $t_c$ decreases, $J_c \propto t^2_c$ decreases more
rapidly. This favors ferromagnetic alignment and deconfinement.
These results are summarized in Fig. 2.

For $\delta$
not too large the hole hopping amplitude $D$ in the isotropic 3-D
case is given by $D_3(\delta) \approx - \delta +
\frac{(6\pi ^2)^{2/3}}{10} \delta ^{5/3}$. In two dimensions,
$D_2(\delta) \approx - \delta + \pi \delta ^2/2$. Hence, the hopping
energy {\em per bond} can be lower in two dimensions. When $t_c$
and $\delta$ are not too small, the system can gain maximum
exchange energy in the $c$ direction and maximum kinetic energy
in the ab plane by having $B_c = D_c = 0$.

To summarize, we have shown that spiral state in the anisotropic
$t$-$J$ model exhibits charge confinement. For a more realistic
treatment one needs to include fluctuations that destroy long-range
magnetic order and reconstruct the Luttinger-Fermi surface, as
shown previously for the 2-D model \cite{sar3}. Such
fluctuations are likely to bring the region of satbility of the
confined state to smaller values of $\delta$ and $t_c/t$.
Nonetheless,
there are some interesting consequences of confinement in our
simple theory. (1) In the confined state the magnetic correlations
along the $c$ direction is peaked at $Q_c = \pi$, while the correlations
in the $ab$ plane remains incommensurate. (2) The $c$-axis resistivity is
strictly infinite at $T = 0$, since <$c^{\dag}_{i\sigma}c_{j\sigma}> =
-D_{ij}B_{ij} = 0$ along the c direction.
 However, at finite $T$ or frequency
there will be an incoherent contribution to the conductivity. Such
a contribution will be activated if there is a spin gap.

\pagebreak

\pagebreak
\centerline {Figure~Captions}

Fig. 1. Phase diagram in the $\delta$-$t_c/t$ plane for two $t/J = 3$
(squares) and $t/J = 5$ (diamonds). In each case, the confined state ($D_c
= 0$) has a lower energy in the V-shaped region.
The transition is first order.

Fig. 2. The hole-hopping amplitude $D_c/(-\delta)$ and the ordering
wavevector along the c-direction Qz $\equiv  \frac {Q_c}{2\pi}$
(diamonds) vs $tc/t$. Note that $Q_c$ is essentially zero
(ferromagnetic) in the unconfined phase. Close to the transition
it aquires a spiral character, and at the transition jumps to
$\pi$ corresponding to antiferromagnetic correlations in the
confined phase.

\end{document}